\documentclass[aps,prl,superscriptaddress,preprintnumbers,
showpacs,a4,twoside,12pt]{revtex4}

\usepackage{bm}
\usepackage{graphicx}
\usepackage{amsfonts}
\usepackage{amsmath}
\usepackage{amssymb}
\usepackage{times}

\usepackage{pdfpages}
\includepdfset{pages=-}

\begin{document}
\title{Bose glass and Mott glass of quasiparticles in a doped quantum magnet}
\author{Rong Yu}
\affiliation{Department of Physics \& Astronomy, Rice University, Houston, TX 77005, USA}
\author{Liang Yin} 
\affiliation{Department of Physics and National High Magnetic Field Laboratory, University of Florida, Gainesville, FL 32611, USA}
\author{Neil S. Sullivan}
\affiliation{Department of Physics and National High Magnetic Field Laboratory, University of Florida, Gainesville, FL 32611, USA}
\author{J. S. Xia} 
\affiliation{Department of Physics and National High Magnetic Field Laboratory, University of Florida, Gainesville, FL 32611, USA}
\author{Chao Huan}
\affiliation{Department of Physics and National High Magnetic Field Laboratory, University of Florida, Gainesville, FL 32611, USA}
\author{Armando Paduan-Filho} 
\affiliation{Instituto de Fisica, Universidade de S\~ao Paulo, 05315-970 S\~ao Paulo, Brasil}  
\author{Nei F. Oliveira Jr.}
\affiliation{Instituto de Fisica, Universidade de S\~ao Paulo, 05315-970 S\~ao Paulo, Brasil}  
\author{Stephan Haas}
\affiliation{Department of Physics and Astronomy, University of Southern 
California, Los Angeles, CA 90089-0484, USA}
\author{Alexander Steppke}
\affiliation{Max-Planck Institute for Chemical Physics of Solids, N\"othnitzer Str. 40, 01187 Dresden, Germany}
\author{Corneliu F. Miclea}
\affiliation{Condensed Matter and Magnet Science, Los Alamos National Lab, Los Alamos, NM 87545}
\author{Franziska Weickert} 
\affiliation{Condensed Matter and Magnet Science, Los Alamos National Lab, Los Alamos, NM 87545}
\author{Roman Movshovich} 
\affiliation{Condensed Matter and Magnet Science, Los Alamos National Lab, Los Alamos, NM 87545}
\author{Eun-Deok Mun} 
\affiliation{Condensed Matter and Magnet Science, Los Alamos National Lab, Los Alamos, NM 87545}
\author{Vivien S. Zapf}
\affiliation{Condensed Matter and Magnet Science, Los Alamos National Lab, Los Alamos, NM 87545}
\author{Tommaso Roscilde}
\affiliation{Laboratoire de Physique, Ecole Normale Sup\'erieure de Lyon,
46 All\'ee d'Italie, 69007 Lyon, France}

\pacs{03.75.Lm, 71.23.Ft, 68.65.Cd, 72.15.Rn}
\begin{abstract} 

{\bf The low-temperature states of bosonic fluids exhibit fundamental quantum 
effects at the macroscopic scale: the best-known examples are  
Bose-Einstein condensation (BEC) and superfluidity, which have been tested
experimentally in a variety of different systems. When bosons are interacting, disorder 
can destroy condensation leading to a so-called Bose glass. This phase has been very 
elusive to experiments due to the absence of any broken symmetry and of a finite energy 
gap in the spectrum. Here we report the observation of a Bose glass of field-induced magnetic
quasiparticles in a doped quantum magnet (Br-doped dichloro-tetrakis-thiourea-Nickel, DTN).
The physics of DTN in a magnetic field is equivalent to that of a lattice gas of bosons
in the grand-canonical ensemble; Br-doping introduces disorder in the hoppings and 
interaction strengths, leading to localization of the bosons into a Bose glass 
down to zero field, where it acquires the nature of an incompressible Mott glass.
The transition from the Bose glass (corresponding to a gapless spin liquid) to the BEC  
(corresponding to a magnetically ordered phase) is marked by a novel, universal exponent 
governing the scaling on the critical temperature with the applied field, in excellent agreement with 
theoretical predictions. Our study represents the first, quantitative account of the universal 
features of disordered bosons in the grand-canonical ensemble.}

\end{abstract}
\maketitle

 \emph{Introduction}. 
  Disorder can have a very strong impact on quantum fluids. Due to their wave-like nature,
  quantum particles are subject to destructive interference when scattering against disordered potentials.
  This leads to their quantum localization (or Anderson localization), which prevents \emph{e.g.} 
  electrons from conducting electrical currents in strongly disordered metals  \cite{KramerM93}, and non-interacting bosons 
  from condensing into a zero-momentum state \cite{Fallanireview}.
  Yet interacting bosons represent a matter wave with arbitrarily strong non-linearity, whose localization 
 properties in a random environment cannot be deduced from the standard theory of Anderson localization.
 For strongly interacting bosons it is known that Anderson localization manifests itself in 
 the \emph{Bose glass}: in this phase the collective modes of the system - and not the individual
 particles - are Anderson-localized over arbitrarily large regions, leading to a gapless energy spectrum, 
 and a finite compressibility of the fluid \cite{GiamarchiS88,Fisheretal89}. 
 Moreover nonlinear bosonic matter waves can undergo a localization-delocalization quantum phase 
 transition in any spatial dimension when the interaction strength is varied
 \cite{GiamarchiS88,Fisheretal89}; the transition brings the system from a non-interacting Anderson insulator 
 to an interacting superfluid condensate, or from a superfluid to a Bose glass. Such a transition is  
relevant for a large variety of physical systems, including superfluid helium in porous 
media \cite{Reppyetal}, Cooper pairs in disordered superconductors \cite{Sacepeetal11}, and cold 
atoms in random optical potentials \cite{Fallanireview,Sanchezpalenciareview}. Despite the long 
activity on the subject, 
a quantitative understanding of the phase diagram of disordered and interacting bosons based on 
experiments is still lacking. 

  Recent experiments have demonstrated the capability of 
realizing and controlling novel Bose fluids made of \emph{quasiparticles} in condensed matter 
systems \cite{Dengetal10, Giamarchietal07}. In this context, a prominent place
is occupied by the equilibrium Bose fluid realized in quantum magnets subject to a 
magnetic field \cite{Giamarchietal07}, in which disorder can be introduced in a controlled
way by chemical doping, leading to novel bosonic phases \cite{disorderedmagnets, 
dopedHpip}. The ground state of such systems without disorder and in zero field 
corresponds to a gapped bosonic Mott insulator. Extra bosons can be injected into the 
system by applying a critical magnetic field that overcomes the gap, and that drives 
a transition to a superfluid state (magnetic Bose-Einstein condensate - BEC).  Such a state 
corresponds to an XY antiferromagnetic state of the spin components transverse to the field. 
  Here we investigate the Bose fluid of magnetic quasiparticles realized in the model 
 $S=1$ compound NiCl$_2$$\cdot$4SC(NH$_2$)$_2$ (dichloro-tetrakis-thiourea-Nickel, DTN) 
 \cite{Zapfetal06}  via experiments (AC magnetic susceptibility, DC magnetization and specific heat), 
 and large-scale quantum Monte Carlo (QMC) simulations. Disorder is introduced by Cl$\to$Br substitution, 
 which, as we will see, leads to randomness in the bosonic hoppings and interactions. 
 We observe a Bose glass in two extended regions of the temperature-magnetic field phase diagram 
 of Br-doped DTN. The gapless nature of the Bose glass manifests itself in a finite uniform magnetic 
 susceptibility (corresponding to the compressibility of the quasiparticles), and in a non-exponential
 decay of the specific heat at low temperature, probing the low-energy density of states. 
 This state extends down to zero field: in this limit the compressibility/susceptibility vanishes,
 while the spectrum remains gapless, giving rise to a \emph{Mott glass}. We investigate
 the thermodynamic signatures of the Mott and Bose glasses, and the Bose-glass-to-superfluid
 transition, characterized by a novel universal exponent for the scaling of the 
 condensation temperature with applied field.  

\begin{figure}[h!]
\begin{center}
\includegraphics[
     width=100mm,angle=0]{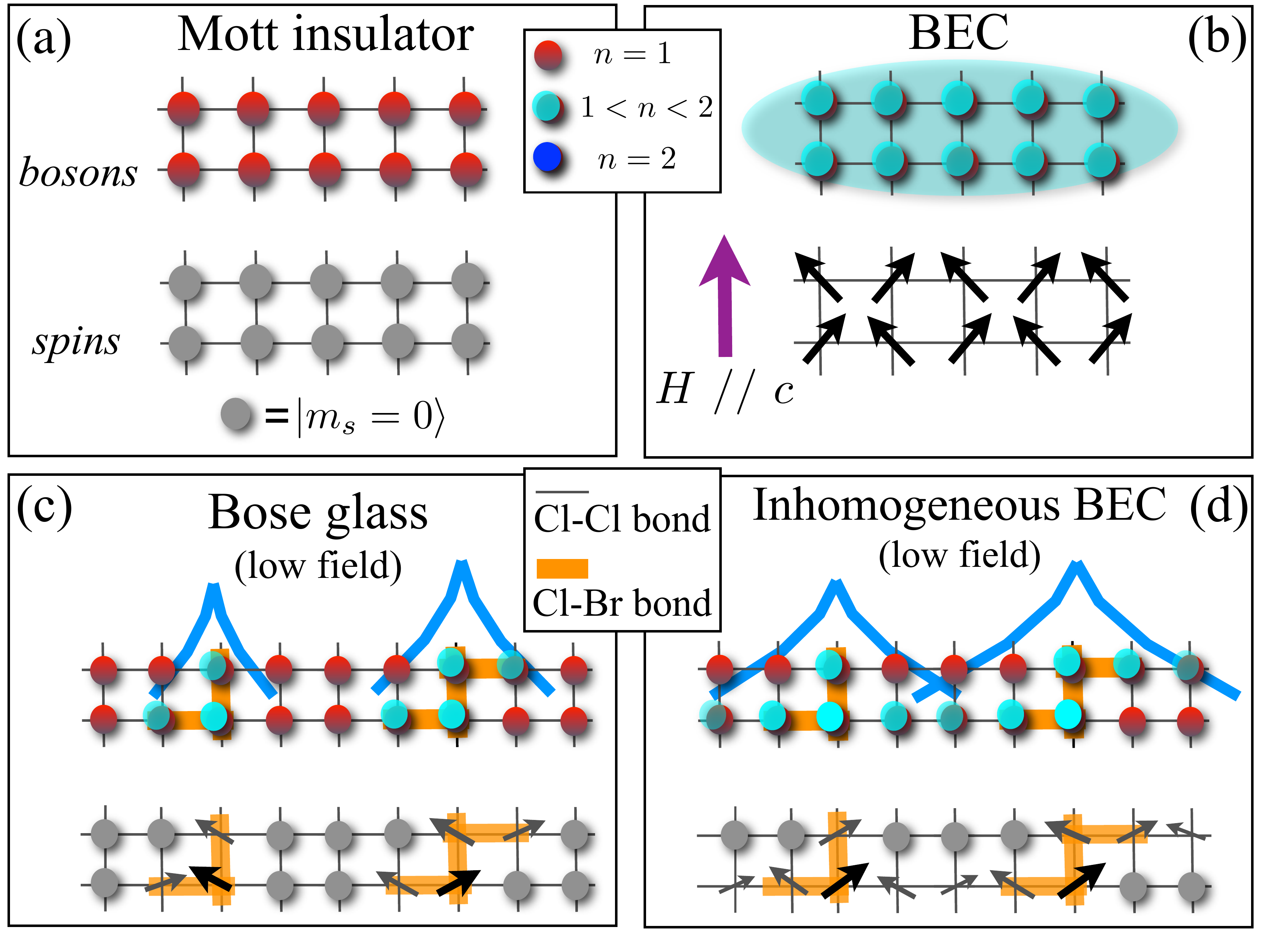} \\
\includegraphics[
     width=100mm,angle=0]{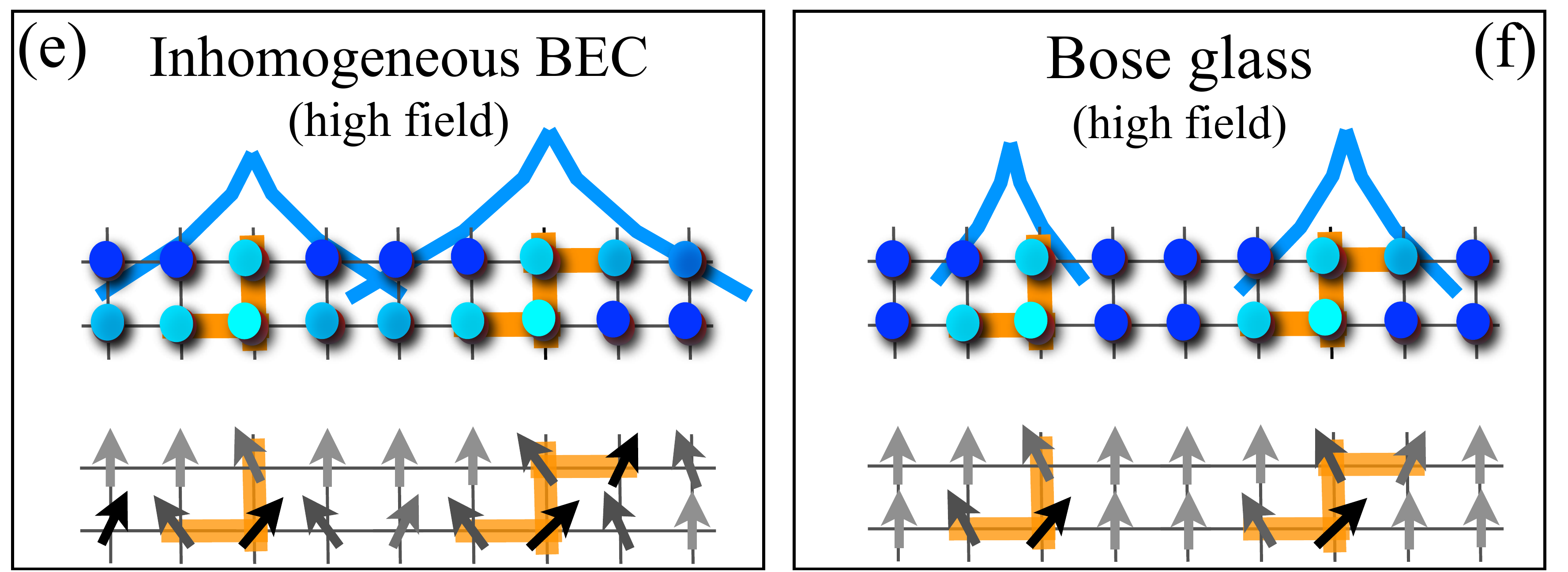} \\     
\caption{Sketch of the bosonic phases of DTN and Br-doped DTN: In the undoped case, an increasing magnetic field along the $c$-axis
drives the system from a Mott insulating (MI) phase (a) to a BEC phase (b) by injecting delocalized excess bosons (indicated in cyan) on top of the 
MI background at density $n=1$; in the doped case, an arbitrarily weak magnetic field can inject extra bosons in the Br-rich regions
(indicated by the orange bonds) which are localized and incoherent in the (low-field) Bose glass phase (c) -- their localized wavefunction is 
sketched by the light-blue lines; further increasing the 
magnetic field leads to the percolation of phase coherence via coherent tunneling of the excess bosons between the 
localized regions, giving rise to an inhomogeneous BEC (d); for strong magnetic fields $H \lesssim H_{c2}$ the spins away from the Br-bonds are close to 
saturation/double occupancy (represented in dark blue), and unpolarized spins / singly occupied sites, corresponding to bosonic holes, 
only survive in the Br-rich regions (e). These holes localized into disconnected, mutually incoherent states when entering the high-field
Bose glass (f).}
\label{f.phases}
\end{center}
\end{figure}

\emph{Magnetic properties of pure DTN.} The magnetic properties of pure DTN are those of antiferromagnetic $S=1$
chains of Ni$^{2+}$ ions, oriented along the crystallographic $c$-axis, and coupled transversely in the $ab$-plane
\cite{Zapfetal06,Zvyaginetal07,notecrystal}. A strong single-ion anisotropy $D$ is present, with an anisotropy axis $z$ corresponding 
to the $c$-axis.  The magnetic Hamiltonian reads
\begin{eqnarray}
{\cal H} &=& J_{c} \sum_{\langle ij \rangle_{c}}  
{\bm S}_{i}\cdot{\bm S}_{j} ~~+ 
J_{ab} \sum_{\langle lm \rangle_{ab}} 
{\bm S}_{l}\cdot{\bm S}_{m} \nonumber \\
&+& D \sum_{i} (S^z_{i})^2
- g\mu_B H \sum_{i}  S^z_{i}.
\label{e.Ham}
\end{eqnarray}
where $J_{c}=2.2$ K is the antiferromagnetic coupling for bonds
$\langle ij \rangle_{c}$ along the $c$-axis, $J_{ab}=0.18$ K is the coupling for bonds 
$\langle lm \rangle_{ab}$ in the $ab$ plane, and $D=8.9$ K 
is the single-ion anisotropy. $g=2.26$ is the gyromagnetic factor $g$ along the 
$c$-axis. In zero field, the large single-ion anisotropy $D$ forces the system into a  
quantum paramagnetic state with each spin close to its $|m_S=0\rangle $ state. 
Mapping the $S=1$ spin states onto bosonic states with occupation $n = m_S +1$,
the quantum paramagnet corresponds to a Mott insulator of bosons with $n=1$
particles per Ni site, and with a gap $\Delta \approx D - 2J_c - 4 J_{ab} + {\cal O}(J^2_c/D)$
for the addition of an extra boson. A magnetic field exceeding the value 
$H^{(0)}_{c1} = \Delta/g\mu_B\approx 2.1$ T  is able to close the spin gap and to create a finite density
of excess bosons that condense into a magnetic BEC (see Fig.~\ref{f.phases}(a)). 
The appearance of excess bosons translates into a finite magnetization
along the field axis; their long-range phase coherence translates into 
long-range XY antiferromagnetic order transverse to the field. 
Long-range order persists up to a critical condensation temperature $T_c$ which, 
for $H\gtrsim H_{c1}^{(0)}$, scales with the applied field as $T_c \sim |H - H_{c1}^{(0)}|^{\phi}$.
Here $\phi=2/3$, as predicted by mean-field theory for a diluted gas of excess bosons, 
and as measured with very high accuracy down to 1mK \cite{Yinetal08}. When 
the magnetic field is increased further, the spins are brought to saturation for 
$H=H_{c2}^{(0)} = (D + 4J_c + 8 J_{ab})/g \mu_B = 12.6$ T, and the system transitions
from a BEC to a perfect Mott insulator with $n=2$ particles per site. Correspondingly,
the BEC critical temperature vanishes as $T_c \sim |H - H_{c2}^{(0)}|^{\phi}$.
\begin{figure}[h!]
\begin{center}
\includegraphics[
    width=140mm,angle=0]{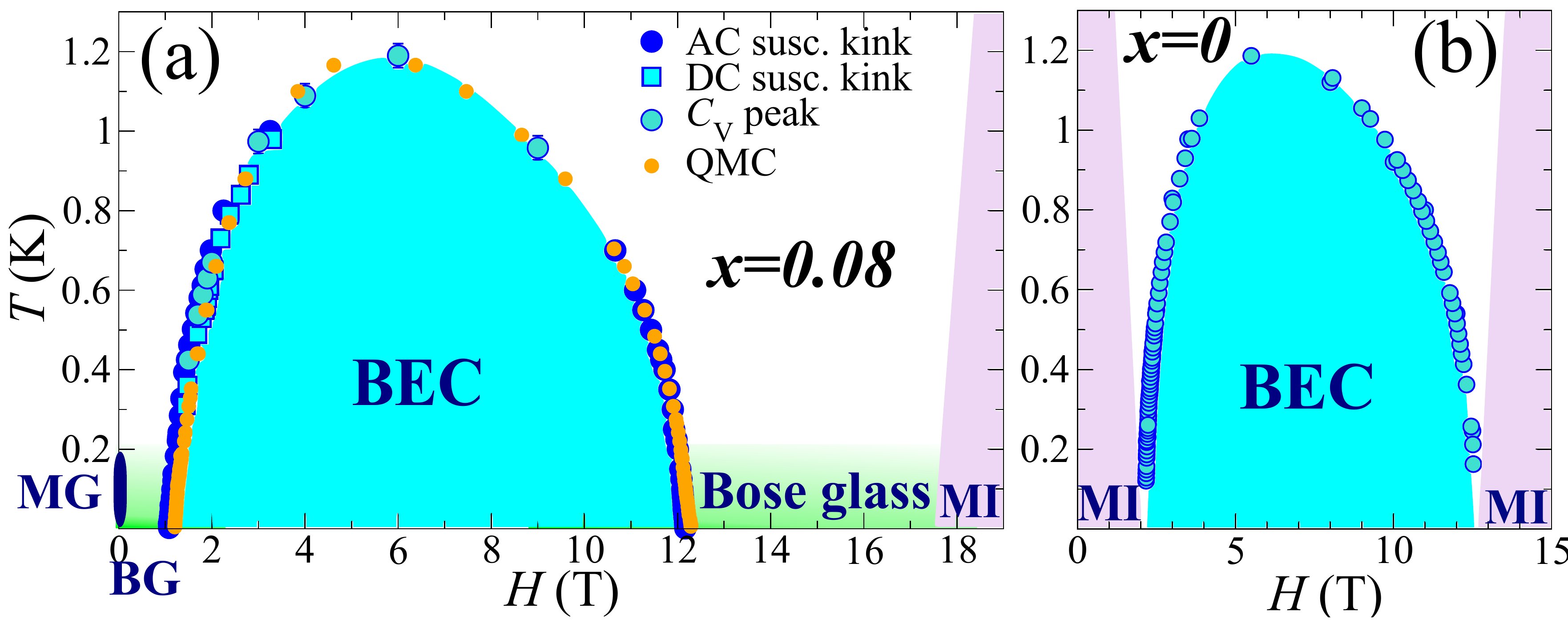} 
\caption{Experimental phase diagram of Br-doped DTN from 
specific heat and susceptometry, compared to QMC data; the following phases are represented:  Bose-Einstein condensation (BEC), 
Bose glass (BG), and Mott glass (MG). The lilac regions represent the magnitude of the spin
gap in the Mott insulating (MI) phase; (b) Experimental phase diagram of 
pure DTN (based on specific heat and the magnetocaloric effect, see Ref.~\cite{Zvyaginetal07}).}
\label{f.phasediagram}
\end{center}
\end{figure}  

\begin{figure}[h]
\begin{center}
\null\hspace*{-.3cm}
\includegraphics[
    width=140mm,angle=0]{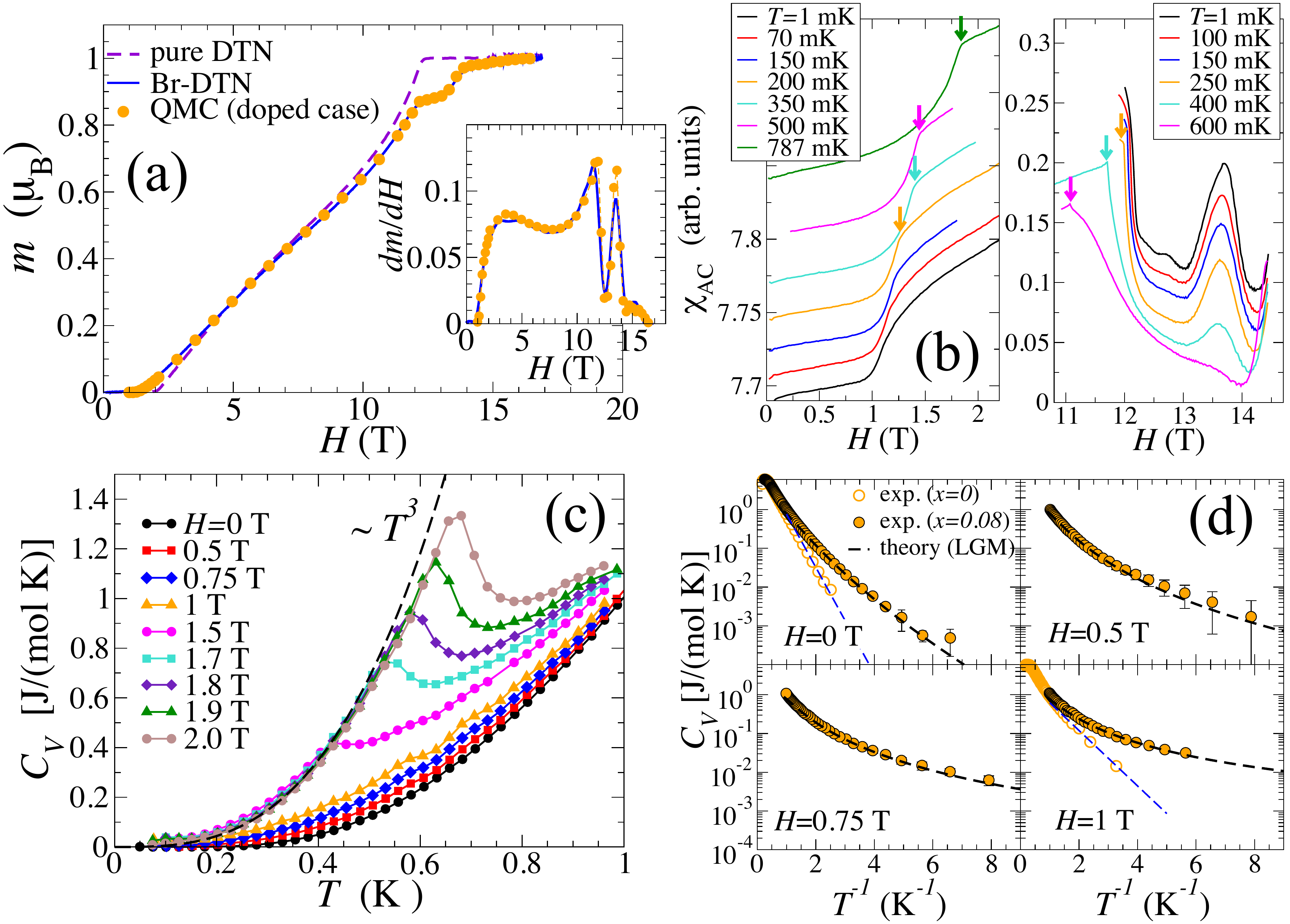} 
\caption{(a) Magnetization curve of Br-DTN at $T=19$ mK, compared to QMC results, and to pure DTN 
magnetization (measured at $T=16$ mK). In the inset we show the DC susceptibility
curve, obtained by differentiating the magnetization;
(b) AC susceptibility of Br-DTN at frequency $f=88.7$ Hz close to the lower and upper critical fields.
The curves have been shifted with respect to one another for readability purposes. The arrows
indicate the appearance of sharp kinks at higher temperatures;
(c) Specific heat of Br-DTN from $H=0$ T to $H=2$ T;
(d) Specific heat of Br-DTN in the MG and BG phase for $H\leq H_{c1} \approx 1$ T, showing a 
non-exponential decay as $T\to 0$; a comparison is made
to the theory predictions based on the local-gap model (LGM), and to the 
data for pure DTN; in the upper-left and lower-right panels, the blue dashed line is a fit of the 
pure-DTN data to $A \exp(-\Delta(H)/k_B T)$ where $A$ is a constant and $\Delta(H)/k_B = g\mu_B (H_{c1}^{(0)}-H)/k_B = 3.16$ K
for $H=0$ and $1.64$ K for $H=1$ T. } 
\label{f.measurements}
\end{center}
\end{figure}

\emph{Phase diagram of Br-doped DTN}. We have measured the critical temperatures and fields for magnetic BEC in 
 Ni(Cl$_{1-x}$Br$_{x})_2\cdot$4SC(NH$_2$)$_2$ (Br-DTN) with $x=0.08\pm 0.005$ doping by measuring the AC susceptibility 
 at low frequencies and the specific
 heat \cite{SuppMat}. AC and DC susceptibility measurements are performed at fixed temperature and varying
 fields, and they show a step-like increase/decrease corresponding to the critical field for BEC, 
 similar to the pure sample \cite{Yinetal08} [see Fig.~\ref{f.measurements}(a)-(b)]. 
 The main difference compared to pure DTN is that -- 
 at low temperatures -- the upper and lower edge of the steps are rounded by disorder;
 as we will see below, this rounding is a fundamental indication of the nature
 of the phases connected by the transition.  An independent estimate of the 
 critical BEC temperature as a function of the field is obtained by the location of a 
 sharp $\lambda$-peak in the specific heat [Fig.~\ref{f.measurements}(c)]. 
 The sharpness of the features corresponding 
 to the BEC transition are quite remarkable, given the strong doping introduced in the system.
 Moreover, for temperatures below the $\lambda$-peak the 
 specific heat clearly follows a $T^3$ behavior, consistent with long-range XY antiferromagnetic
 order in 3D.  
 Fig.~\ref{f.phasediagram} summarizes the experimental phase diagram of Br-DTN. 
 Br-doping has a profound impact on the phase diagram of DTN: in particular \emph{both} the lower
 and upper critical fields for the onset of magnetic BEC at $T\to 0$ are found to shift 
 to lower values, $H_{c1} = 1.07(1)$ T  and $H_{c2} = 12.16(1)$ T, as shown in Fig.~\ref{f.phasediagram}.  
But most importantly the magnetic behavior of Br-DTN outside the BEC region is completely
different compared to the pure system. In the pure system, the ground state outside the magnetic BEC
is a Mott insulator with a large spin gap $\Delta$ away from the critical fields. This leads to an exponential
suppression of the specific heat at low temperatures $k_B T \lesssim \Delta$ as 
$C_V \sim \exp[-\Delta/(k_B T)]$, as shown in Fig.~\ref{f.measurements}(d), and to a similarly vanishing 
susceptibility for $T\to 0$.  On the contrary, for $x=0.08$, we observe that the 
susceptibility is finite for $H\geq H_{c2}$, and it even exhibits a strong satellite peak 
for  $H\approx 13.5$ T. The susceptibility vanishes only for $H = H_s \approx  17$ T, 
corresponding to the saturation field of the entire sample, which is pushed to a much higher
value than in the pure sample (where $H_s = H_{c2}^{(0)} = 12.6$ T).   
In the region $H \leq H_{c1}$ we observe that the specific heat exhibits a 
non-exponential decay, \emph{down to zero field} [Fig.~\ref{f.measurements}(d)]. 
 Therefore we can conclude that the non-magnetic phases for $0< H \leq H_s$ correspond to 
 \emph{gapless} bosonic insulators, which, as we will see, can be identified with
 a compressible \emph{Bose glass} (for $H>0$) and an incompressible \emph{Mott glass}
 (for $H=0$). 

\emph{Modeling Br doping}. Br-DTN can be successfully modeled theoretically by considering that Br substitution for Cl 
 affects the super-exchange paths associated with the $J_c$ couplings, and it also distorts 
 the lattice locally due to the larger atomic radius of Br with respect to Cl. The disappearance of the spin gap down 
 to $H=0$ and the upward shift of the saturation field suggests that Br doping locally strengthens 
 the magnetic coupling $J_c$ and lowers the anisotropy $D$. For simplicity we only 
 consider that Ni-Cl-Cl-Ni bonds in DTN can be turned into Ni-Cl-Br-Ni or Ni-Br-Cl-Ni, 
 and we neglect Ni-Br-Br-Ni bonds that represent only $ 0.6\%$ of the total bonds for $x=0.08$.
 We assign a $J_c'$ value to the magnetic exchange coupling of the Br-doped bonds, 
 and a $D'$ value to the single-ion
anisotropies of the Ni ion adjacent to the Br dopant. Note that for a doping concentration $x$, we 
have a fraction of $2x$ doped bonds, given that each bond can accommodate a Br dopant
on two different Cl sites.  We then use $J_c'$ and $D'$ as fitting
parameters of the full low-temperature magnetization curve in Fig.~\ref{f.measurements}(a), 
which is calculated using QMC simulations \cite{SuppMat}.  
We find an extremely good agreement between experimental data and simulation for 
$J_c' \approx 2.35 J_c$ and $D' \approx D/2$, giving us confidence that we are able
to quantitatively model the fundamental microscopic effects of doping in Br-DTN.  
Indeed the critical temperature for BEC, extracted from a finite-size scaling analysis of 
the simulation data with doping $x=0.075$ \cite{SuppMat}, is in remarkable quantitative agreement 
with the experiment, as shown in Fig.~\ref{f.phasediagram}(a). The critical fields estimated  
from simulations are $H_{c1} =  1.199(5)$ T and  $H_{c2} = 12.302(5)$ T, slightly larger (by 
$\sim 0.13-0.14$ T)
than the experimental values. However the downward shifts of $H_{c1}$ and
$H_{c2}$ with respect to the pure system are correctly captured. 
  
 \emph{Bose and Mott glass.} The anisotropy $D$ acts as a repulsion term inhibiting two bosons 
 from sitting on the same site, while the coupling $J_c$ controls the kinetic
 energy of the bosons. In particular we find numerically \cite{SuppMat} that a model in which
 all $c$-axis bonds contain a Br dopant (leading to a couplings $J_c'$ and to an anisotropy
 $D'$ on one of the two sites connected by the bond), is in a BEC phase (XY ordered
 phase) even in zero field. This means that the Br-rich regions in DTN, characterized
 by the Hamiltonian parameters $J_c'$ and $D'$,  behave locally as mini-BECs, and they
 are \emph{locally gapless}. 
 Strictly speaking, Br-rich regions will have a residual gap due to their finite size.
 However, the statistical distribution of sizes has no upper bound, so that the
 distribution of local gaps has no lower bound, and consequently 
 Br-DTN is globally gapless even in zero field. The corresponding bosonic phase
 is therefore a gapless insulator with 
 spin inversion symmetry along field axis, and resulting in a commensurate
 boson density $n=1$. This represents to our knowledge the first experimental
 realization of a \emph{Mott glass} \cite{ProkofevS04, RoscildeH07}, which 
 has the peculiar aspect of being incompressible (namely of having a vanishing 
 susceptibility at $T=0$) despite being gapless \cite{RoscildeH07, SuppMat}. 
 As soon as a field is applied to the system, it can immediately inject excess bosons,
 which localize \emph{\`a la Anderson} in the Br-rich regions, resulting 
 in a paradigmatic example of a Bose glass (Fig.~\ref{f.phases}(c)). In the spin 
 language, spins in the Br-rich regions acquire a finite magnetization along the 
 field and their transverse components correlate antiferromagnetically over
 a finite range, but the local phase of the antiferromagnetic order is different
 from region to region so that the system remains globally paramagnetic. 
Long-range phase coherence of the local order parameters - corresponding to 
the local phases of the bosonic wavefunction - is established only when the localized
states of the bosons grow enough under the action of the applied field as to overlap, 
leading to coherent tunneling of bosons between neighboring localized states (Fig.~\ref{f.phases}(d)).
The resulting phase is a highly inhomogeneous BEC \cite{Yuetal10}.   
 
\begin{figure}[h]
\begin{center}
\includegraphics[
  width=100mm,angle=0]{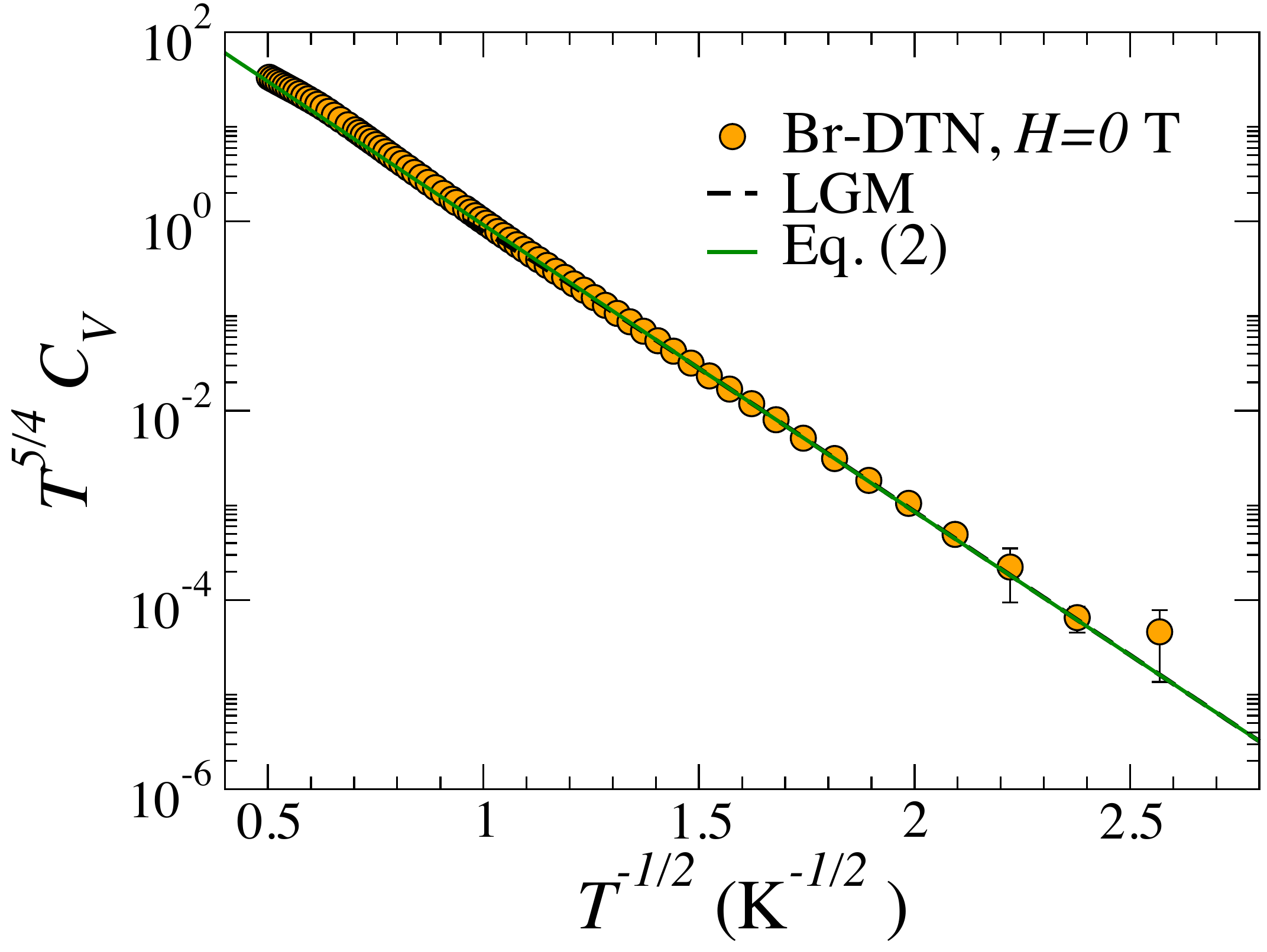} 
\caption{Specific heat in zero field, displaying the characteristic Mott glass
scaling. The solid lines are theoretical predictions based on the LGM, with
parameter $c=3.02$.}
\label{f.cv}
\end{center}
\end{figure}

 We can quantitatively test the picture of bosons localized in Br-rich regions against
 the thermodynamic behavior of Br-DTN by using a simplified \emph{local-gap model}
 (LGM). Within this model \cite{SuppMat}, the low-temperature and low-field behavior of the system 
 is reduced to that of a collection of three-level systems, corresponding to a local 
 longitudinal magnetization $m_{S,{\rm tot}} = 0, \pm 1$ for each localized state.  
 There is a finite-size gap $\Delta_N \approx c/N$ (for zero field) between the $m_{S,{\rm tot}} = 0$ 
 ground state and the $m_{S,{\rm tot}} = \pm 1$ excited states, where $N$ is the number of 
 sites in the Br-rich cluster. The low-temperature specific heat in zero field can then be 
 predicted analytically to be 
 \begin{equation}
 C_V(T) \sim t^{-5/4}  \exp\left(-2\sqrt{cx_0/t}\right)
 \label{e.cv}
 \end{equation}
 where $t = k_B T/J_c$ and $x_0 = \log(2x)$; this expression displays the 
 stretched exponential behavior, that uniquely characterizes the Mott glass \cite{RoscildeH07}.
 The $c$ parameter, and an overall prefactor, are used as fitting parameters of the 
 experimental data in zero field, leading to an extremely good fit, as shown in Figs.~\ref{f.measurements}(d) and \ref{f.cv}. 
Notably, no further adjustable parameters are necessary to fit the finite-field data, 
displayed in Fig.~\ref{f.measurements}(d), which also show a remarkable agreement
with the theory prediction up to $H\approx H_{c1}$. 

For $H\to H_{c2}$ the magnetization approaches the value $m_x = 1-2x \approx 0.84$, where 
all spins not connected to a Br-doped bond are polarized -- and indeed $H_{c2}$
lies very close to the polarization field $H_{c2}^{(0)}$ of pure DTN. The full polarization of the Br-poor
regions leads to a \emph{pseudo}-plateau in the magnetization at $m\approx m_x$ (\emph{pseudo} because it 
still exhibits a small finite slope). This feature corresponds to the high-field
BG phase, which is characterized by the localization of bosonic \emph{holes},
or singly occupied sites with $m_S=0$, in a background of doubly occupied sites with $m_S=1$ (Fig.~\ref{f.phases}(f)). 
Such holes persist up to the saturation field $H_s$, which is the field necessary to
fully polarize a homogeneous system with $J_c'$ couplings and $D'$ anisotropies everywhere. 
The step-like feature in the magnetization at the upper bound of the pseudo-plateau
is therefore induced by the 
saturation of the Br-rich clusters, and it is smeared due to the fact that such clusters 
have random geometries and therefore a distribution of local saturation fields, 
upper bounded by $H_s$. 
 
 \begin{figure}[h!]
\begin{center}
\includegraphics[
  width=140mm,angle=0]{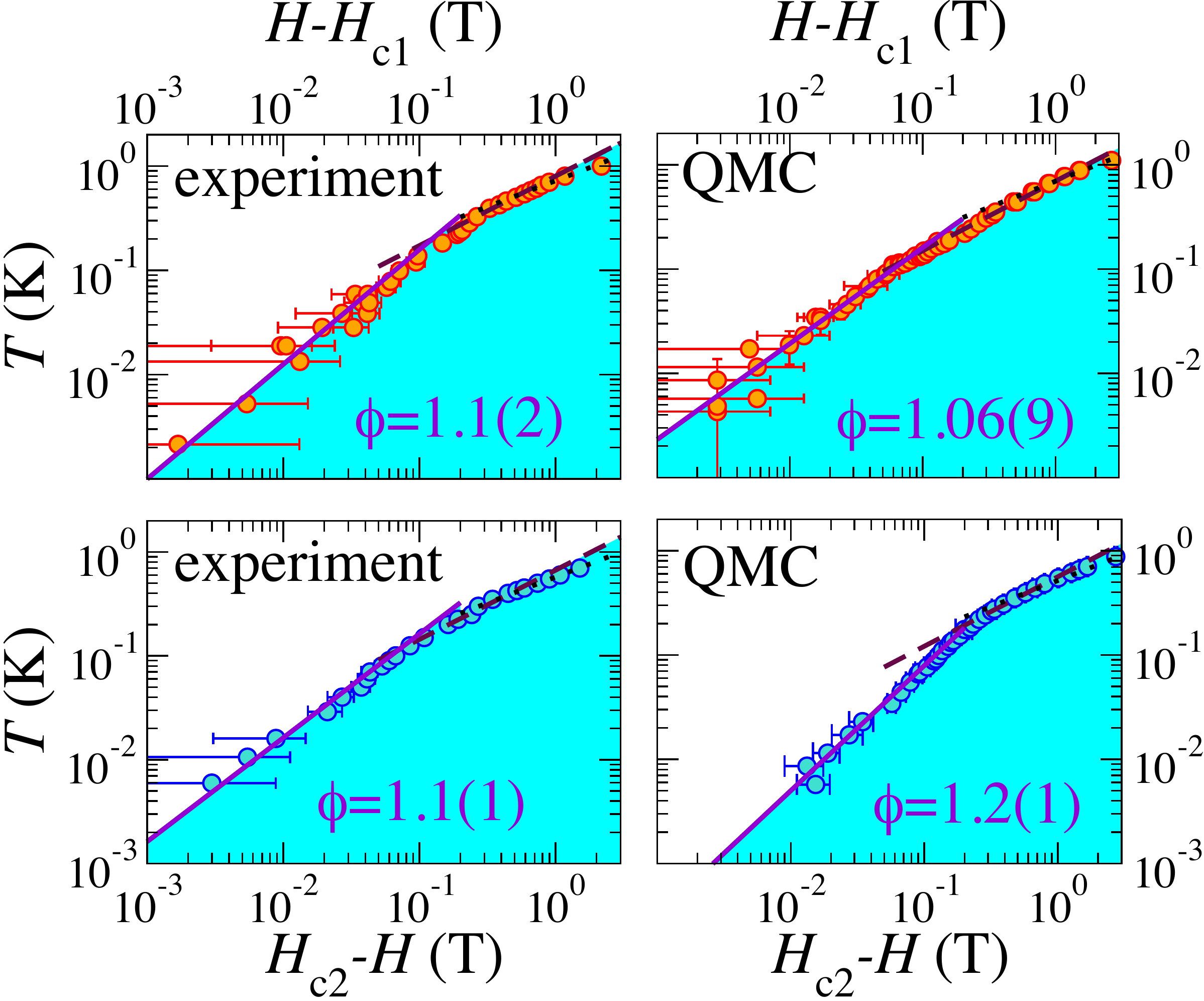} 
\caption{Scaling of the critical temperature with the distance from the $T=0$ critical fields, 
exhibiting a crossover between various exponents. The dashed and dotted lines indicate a fit to 
the form $a |H - H_{c1(2)}| ^{2/3}$ and $a |H - H_{c1(2)}| ^{1/2}$ respectively, while the solid line is a fit to 
$a' |H - H_{c1(2)}| ^{\phi}$, with the resulting $\phi$ exponent indicated
in the figure ($a$ and $a'$ are fitting parameters). The leftmost panels show the critical line 
extracted from the AC susceptibility, and the rightmost ones the critical line obtained
from the QMC simulations.}
\label{f.crossover}
\end{center}
\end{figure}  
 
\emph{Thermal percolation crossover.} 
The physics described so far is valid only for low temperatures. As the temperature is 
increased above $\sim 200$ mK, the bosons that were localized in the Bose Glass state thermally 
delocalize and proliferate; this leads to a thermal percolation of their density profile
(corresponding to the longitudinal magnetization profile) throughout the sample  \cite{Yuetal10}. 
Thus a more ordinary paramagnet forms at higher 
temperatures and the nature of field-driven transition into the BEC phase also changes fundamentally.
Indeed at temperatures 
below the thermal percolation crossover, the BEC transition occurs as sketched in 
Figs.~\ref{f.phases}(c)-(d) and Figs.~\ref{f.phases}(e)-(f), namely by coherent tunneling 
of bosons between localized states, resulting in a highly inhomogeneous BEC phase. This picture
changes above the thermal percolation crossover. Now in the normal phase, the 
bosons move incoherently on a pre-percolated network of magnetized sites, and 
their BEC transition upon increasing the field corresponds therefore to condensation 
on a random 3D percolated lattice, which is fully analogous to condensation
on a regular 3D lattice. 
Signatures of the thermal percolation crossover can be found in the critical behavior of the 
AC susceptibility: At low temperatures ($T \lesssim 200$ mK) it exhibits a rounded shoulder for 
$H \gtrsim H_{c1}$ and $H \lesssim H_{c2}$, and at higher temperatures it shows a sharp kink --
 analogous to what is observed in the pure system \cite{Yinetal08} 
 (see Fig.~\ref{f.measurements}(b)). \\
 But the most dramatic signature of the thermal percolation crossover is observed in the scaling 
 of the critical temperature with the applied field, shown in Fig.~\ref{f.crossover}. 
 Plotting $T_c$ vs. $|H - H_{c1(2)}|$ on a log-log scale, we clearly observe  
 a kink separating two different scaling regimes. At high temperatures ($T \gtrsim 200-300$ mK) 
 the field-dependence of $T_c$ is essentially consistent with a pure-system scaling for low
 temperatures, 
 $T_c \sim |H - H_{c1(2)}|^\phi$ with $\phi = 2/3$, or with a pure-system scaling 
 for intermediate temperatures with $\phi=1/2$, as observed in other magnetic
 BEC systems \cite{Kawashima04} . At low temperatures, the scaling exponent
 crosses over to novel values, $\phi = 1.1(2)$  (close to $H_{c1}$) and 
 $\phi = 1.1(1)$ (close to $H_{c2}$), which are consistent within the error
 (see \cite{SuppMat} for a discussion of the estimate of $\phi$). 
 Moreover, these novel scaling exponents are consistent as well with the values extracted
 from our QMC simulations ($\phi = 1.06(9)$ and $1.2(1)$ close to $H_{c1}$ and $H_{c2}$ 
 respectively). Simulations also show a rough quantitative agreement for the crossover 
temperature range. Most remarkably, a consistent value
of the exponent $\phi$ at low temperature is also observed theoretically for the magnetic Hamiltonian of DTN
subject to a different type of disorder, namely site dilution \cite{Yuetal10}. We can therefore 
conclude that the low-temperature scaling of $T_c$ exhibits a novel exponent $\phi \sim 1-1.1$
which is a \emph{universal} feature of the BG-BEC transition. 

 \emph{Conclusions.} We have performed a comprehensive experimental and theoretical study
 of the disordered and strongly interacting Bose fluid realized in a doped quantum magnet
  (Br-DTN) under application of a magnetic field. We provide substantial evidence of the existence of
 gapless insulating phases of the bosons - the Mott glass and the Bose glass - and we
 investigate for the first time the quantitative features associated with their thermodynamic
 behavior. These phases can be quantitatively described as a Bose fluid
 fragmented over an extensive number of localized states with variable local gaps, 
 dominating the response of the system. 
 The presence of a Bose glass leads to a novel and seemingly universal exponent governing 
 the scaling of the critical temperature for the transition from Bose glass to BEC.   
 The remarkable agreement between theory and experiment shows that Br-DTN 
 is an extremely well controlled realization of a disordered Bose fluid, which
 allows a detailed experimental study of the thermal phase diagram of disordered bosons
 in the grand-canonical ensemble. 
 
\emph{Acknowledgements.}
 Work at the High Magnetic Field Laboratory at the Physics Institute of the University of Sao Paulo 
 were supported in part by the Brazilian agencies FAPESP and CNPq.
 Measurements at the NHMFL High B/T and pulsed field facilities were supported by NSF Grant DMR 0654118, 
 by the State of Florida, and the DOE.
 Work at LANL was supported by the NSF, and the DOE's Laboratory Directed Research and Development 
 program under 20100043DR. The numerical simulations have been performed on
the computer facilities of the NCCS  at the Oak Ridge National Laboratories, 
and supported by the INCITE Award MAT013 of the Office of Science - DOE.



\section*{Supplementary material}

Rong Yu,[1] Liang Yin,[2] Neil S. Sullivan,[2] J. S. Xia,[2] Chao Huan,[2] Armando Paduan-Filho,[3]
Nei F. Oliveira Jr.,[3] Stephan Haas,[4] Alexander Steppke,[5] Corneliu F. Miclea,[6] Franziska
Weickert,[6] Roman Movshovich,[6] Eun-Deok Mun,[6] Vivien S. Zapf,[6] and Tommaso Roscilde[7]

[1]*Department of Physics & Astronomy, Rice University, Houston, TX 77005, USA*
[2]*Department of Physics and National High Magnetic Field Laboratory, University of Florida, Gainesville, FL 32611, USA*
[3]*Instituto de Física, Universidade de São Paulo, 05315-970 São Paulo, Brasil*
[4]*Department of Physics and Astronomy, University of Southern California, Los Angeles, CA 90089-0484, USA*
[5]*Max-Planck Institute for Chemical Physics of Solids, Nöthnitzer Str. 40, 01187 Dresden, Germany*
[6]*Condensed Matter and Magnet Science, Los Alamos National Lab, Los Alamos, NM 87545*
[7]*Laboratoire de Physique, Ecole Normale Supérieure de Lyon, 46 Allée d'Italie, 69007 Lyon, France*

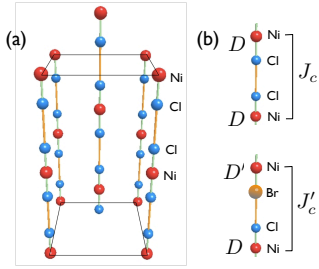

FIG. 1: (a) NiCl$_2$ lattice in DTN; (b) effect of Br doping on the super-exchange coupling between Ni$^{2+}$ ions.

## SAMPLE PREPARATION AND EXPERIMENTAL METHODS

DTN crystals were prepared by evaporation of an aqueous solutions of thiourea and nickel chloride. A slight excess of hydrochloric acid facilitates the growth of the samples. The solution for the doped crystals were prepared by thiourea, nickel chloride and hydrobromic acid. The solution temperature was kept around 35 degrees centigrade. The amount of Br/Cl atoms in the sample were determined by potentiometric analysis. The structure of the Ni-Cl lattice in DTN is shown in Fig. 1(a). The Ni atoms form two interpenetrating tetragonal systems which are effectively decoupled from a magnetic point of view [1]. Br substitution alters the bonds connecting two adjacent Ni atoms, and it is found to alter the magnetic coupling between the Ni ions (from $J_c$ to $J'_c$) as well as the on-site single-ion anisotropies (from $D$ to $D'$).

All magnetic measurements were made with the magnetic field applied along the tetragonal axis ($c$-axis) of the sample. The AC susceptibility measurements were carried out using a PrNi$_5$ nuclear refrigerator (down to 1 mK) and a 15 T magnet at the High B/T facility of National High Magnetic Field Laboratory (NHMFL). The experimental methods have been described elsewhere [2]. The AC susceptibility measurements allow to extract the critical field for magnetic BEC at fixed temperature, $H = H_c(T)$, which provides the magnetic-field dependence of the critical temperature by inversion, $T = T_c(H)$. The lower/upper critical field $H_c$ is determined by a peak/dip in the first derivative of the AC susceptibility with respect to the applied field.

The DC magnetization was measured at the Physics Institute of the University of Sao Paulo by using a plastic dilution refrigerator at 19mK. The main magnetic field was generated by a Nb$_3$Sn superconducting magnet, with a maximum field of 17 T. The field sweep rate, [dB/dt], was 0.43 T/min. The magnetization was measured by the force method: the magnetic force acting on the sample was produced by a magnetic-field gradient, generated by a superconducting gradient coil superimposed on the main magnetic field, and it was measured by a capacitance technique. Specific heat was measured in a Quantum Design $^3$He/$^4$He dilution refrigerator down to 50 mK using the thermal relaxation method.

## THEORETICAL METHODS

### Spin-to-boson mapping

The model for the magnetic behavior of DTN and Br-doped DTN, Eq. (1) of the main text, can be exactly cast in the form of a Hamiltonian for interacting bosons via the Holstein-Primakoff spin-boson transformation for $S = 1$ spins: $S^- = \sqrt{2}\sqrt{1 - n/2}\, b$, $S^+ = \sqrt{2}\, b^\dagger \sqrt{1 - n/2}$, and $S^z = n - 1$. Here $n = b^\dagger b$, and $b, b^\dagger$ are triplet-boson operators satisfying the commutation relation $[b, b^\dagger] = 1$ and the constraint $n = \{0, 1, 2\}$. The magnetic Hamiltonian modeling Br-doped

DTN takes then the bosonic form

$$
\begin{aligned}
\mathcal{H} = & -\sum_{\langle ij \rangle} J_{ij} \left[ \sqrt{1 - \frac{n_i}{2}} \, b_i b_j^\dagger \sqrt{1 - \frac{n_j}{2}} + \text{h.c.} \right] \\
& + \sum_{\langle ij \rangle} J_{ij}(n_i - 1)(n_j - 1) + \sum_i D_i(n_i - 1)^2 \\
& - g\mu_B H \sum_i n_i + \text{const.}
\end{aligned}
\tag{1}
$$

Here $J_{ij} = J_{ab}$ for nearest-neighbor bonds $\langle ij \rangle$ lying in the $ab$ plane; for bonds along the $c$ axis, $J_{ij}$ can take values $J_c$ or $J_c' = 2.3 J_c$ with probabilities $1 - 2x$ and $2x$ respectively, where $x$ is the Br concentration. The single-ion anisotropy $D_i$ takes the value $D'$ when the site is adjacent to a Br dopant, and value $D$ otherwise (with $D' = D/2$).

It is evident that Eq. (1) has the form of an extended Bose-Hubbard Hamiltonian describing nearest neighbor hopping, on-site and nearest neighbor interactions of bosons on a lattice. In particular, the magnetic couplings play the role of an hopping term for the bosons as well as of nearest neighbor repulsion. In Eq. (1) we have exploited the bipartite nature of the magnetic lattice to change the sign of the hopping term via a gauge transformation $b_i \to -b_i$ on a sublattice. Compared to a standard Bose-Hubbard Hamiltonian, extra $\sqrt{1 - \frac{n_i}{2}}$ factors appear in the hopping term, which provide the correct values of the matrix elements of the spin operators $S^+$ and $S^-$, and in particular they constrain the bosonic occupation number. They do not alter the U(1) symmetry of the Hamiltonian, which is spontaneously broken by the BEC state (XY ordered state). Therefore Eq. (1) shares the same universal critical features of a standard Bose-Hubbard Hamiltonian. The single-ion anisotropy plays the role of an on-site repulsion which is particle-hole symmetric around the $n = 1$ state, namely it penalizes equally the double occupancy $n = 2$ and the vacuum $n = 0$. The nearest-neighbor repulsion due to $J_{ij}$ is also particle-hole symmetric, and it favors a doubly occupied site neighboring an empty one. However the dominant $D$ anisotropy makes this configuration very unlikely, so that the n.n. repulsion plays a marginal role. Br-DTN can thus be modeled using an extended Bose-Hubbard Hamiltonian with random hoppings, random on-site interactions, and random nearest-neighbor interactions, which are bimodally distributed in a spatially correlated way.

## Quantum Monte Carlo simulations

To obtain the theoretical phase diagram of the model for Br doped DTN, we perform numerically exact quantum Monte Carlo (QMC) simulations based on the stochastic series expansion algorithm [3]. The simulations are done on cubic lattices with up to $18^3$ quantum spins and the simulation results are averaged over 300 disorder realizations. For some selected temperatures and fields we have run simulations on lattice with $20^3$ spins and averaging over 1200 disorder realizations. The critical temperatures and critical fields have

been estimated via a careful finite-size scaling analysis of the spin correlation length, which is calculated via the disorder-averaged second-moment estimator [4]:

$$
\xi_p = \frac{L}{2\pi} \sqrt{\frac{[S(\boldsymbol{Q})]_{av}}{[S(\boldsymbol{Q} + \frac{2\pi}{L}\hat{x}_p)]_{av}} - 1}.
\tag{2}
$$

Here $L$ is the linear dimension of the finite system, $\hat{x}_p$ refers to the unit vector along the $p$ ($p = a(b)$, $c$) crystal axis, and the ordering wavevector $\boldsymbol{Q} = (\pi, \pi, \pi)$. $[\ldots]_{av}$ denotes the disorder average. $S(\boldsymbol{Q})$ is the static spin structure factor for the transverse spin components,

$$
S(\boldsymbol{Q}) = \frac{1}{L^3} \sum_{i,j} e^{i\boldsymbol{Q}\cdot(\boldsymbol{r}_i - \boldsymbol{r}_j)} \langle S_i^{x(y)} S_j^{x(y)} \rangle .
\tag{3}
$$

Near the critical point, the correlation length follows the scaling ansatz

$$
\xi_p/L \approx \mathcal{F}_{\xi_p}^{1(2)} [(h - h_{c1(2)}) L^{1/\nu}],
\tag{4}
$$

which is verified by the QMC data shown in Fig. 2. At each temperature, we determine the critical field $h_{c1(2)}$ as the crossing point of the $\xi/L$ curves for different lattice sizes.

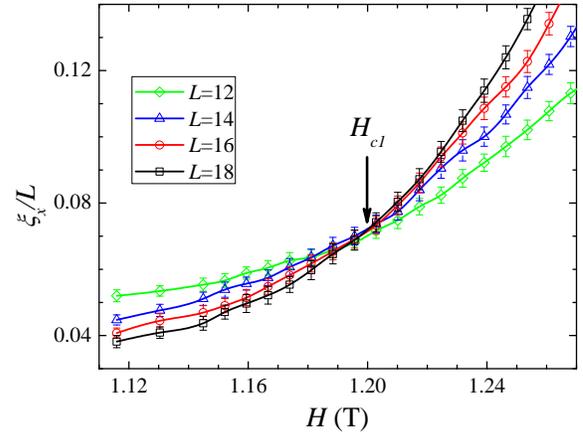

FIG. 2: Finite-size scaling of the correlation length at $T = 172$ mK in the Br-doped DTN model with $x = 0.075$.

We also applied the QMC simulations to study the phase diagram of the system with homogeneous couplings $J_c'$ on all $c$-axis bonds and having anisotropies $D'$ on one of the two sites connected by the bond. This models the idealized situation (denoted as Br-only DTN) in which all all the bonds along the $c$-axis are of the Ni-Br-Cl-Ni type, and in which the position of the $Br$ atom is randomly chosen between the two Cl sites; therefore the anisotropies are randomized, taking on average the value $D'$ on 3/4 of the sites and $D$ on the remaning 1/4. Although not realizable in the experiments as such,

this model extends the behavior of the Br-rich regions in real Br-doped DTN to the thermodynamic limit, and therefore it unveils the local response of the Br-rich regions. Following a similar analysis of the correlation length as that described above, we find that Br-only DTN displays XY antiferromagnetic order up to a finite critical temperature even in *zero* magnetic field, and it therefore exhibits a gapless spectrum. As shown in Fig. 3, the finite-size scaling of $\xi/L$ is consistent with a divergence in the thermodynamic limit at $H = 0$ and a finite temperature $T = 68.8$ mK, indicating that the system is long-range ordered at this temperature. We notice that this is a non-trivial result, given that we are dealing with a random-$D$ system. On the other hand we have studied separately the non-random situation in which all anisotropies take the value $D$, finding that even in this case the system displays long-range order – and *a fortiori* this is verified when all anisotropies take the value $D' < D$. Therefore Br-only DTN realizes a random arrangement of local environments (with $D$ anisotropy or $D'$ anisotropy) which are both gapless and ordered in the thermodynamic limit.

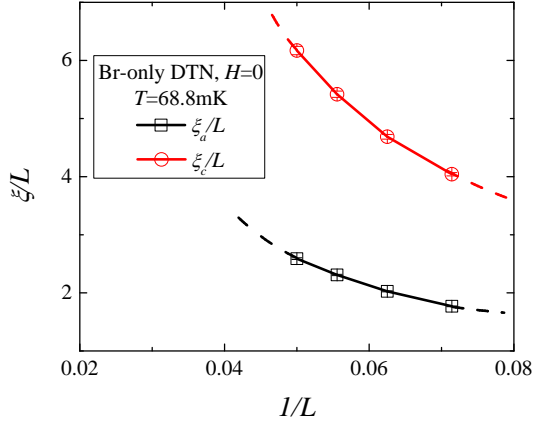

FIG. 3: Finite-size scaling of the correlation length at zero magnetic field and at $T = 68.8$ mK in the Br-only DTN model with $J'_c = 2.35 J_c$ and $D' = 2J_c$ everywhere.

The gapless nature of the spectrum of Br-only DTN in zero magnetic field is inherited by Br-doped DTN, which remains gapless down to zero magnetic field regardless of the doping concentration. This is due to the fact that Br-rich regions, locally mimicking Br-only DTN, can have arbitrarily large sizes and therefore arbitrarily small local gaps. The Br-rich regions, although exhibiting local correlations similar to Br-only DTN, are not correlated with each other, so that the system in zero field does not possess long-range order, resulting in a Mott glass state.

## Local gap model

The local gap model (LGM) describes a statistical ensemble of Br-rich clusters, that are gapless in the thermodynamic limit even in zero field. Exact diagonalization shows that small one-dimensional Br-rich clusters, modeled by the anisotropic Heisenberg model

$$\mathcal{H}_{\text{Br}} = J'_c \sum_i \boldsymbol{S}_i \cdot \boldsymbol{S}_{i+1} + D' \sum_i (S_i^z)^2 \quad (5)$$

(with $D'/J'_c = 0.86$) have a spectrum with an $m_S = 0$ ground state, and first excited states with $m_S = \pm 1$ at an energy difference $\Delta_N$ (where $N$ is the size of the cluster), vanishing in the thermodynamic limit. Therefore the application of a magnetic field to the system leads to a splitting of the first two excited states, leaving the ground state unaffected. The energy of the two excited states become then

$$\Delta_N^{(\pm)} = (\Delta_N \mp h) \, \theta(\Delta_N \mp h) \quad (6)$$

where $h = g\mu_B H/J_c$. Here $\Delta_N$ is expressed in units of $J_c$. The $\theta$-function term is introduced *ad hoc* so as to make the cluster gapless when the local gap is closed by the field, and it prevents a gap from re-opening when $h > \Delta_N$. In fact a re-opening of the gap is not physically motivated, given that an increase of the field will make the size of the locally magnetized region increase, thereby increasing the local density of states at low energy and not the opposite. For a cluster of size $N$, we take the following scaling ansatz for the gap in a finite applied field

$$\Delta_N \approx \frac{c}{N} \,, \quad (7)$$

which is typical of Heisenberg antiferromagnets undergoing spontaneous symmetry breaking [5]. Nonetheless, for irregularly shaped clusters this is obviously a strongly simplifying assumption, given that the above Ansatz only depends on the size of the cluster, and not on its geometry.

At low temperatures, $t = k_B T/J_c \lesssim \Delta_{\langle N \rangle}$ with $\langle N \rangle \approx 1 + 1/x_0$, where $x_0 = |\log(2x)|$ we can restrict the spectrum to the three low-energy states depicted above.

Therefore the thermodynamics of the system reduces to that of a three-level system, with the following formulas for the thermal energy $\Delta E_N(T) = E_N(T) - E_N(0)$ and total magnetization $M_N(T)$ of a cluster of size $N$:

$$M_N(T) = \frac{1}{Z_N} \left[ e^{-\beta \Delta_N^{(+)}} - e^{-\beta \Delta_N^{(-)}} \right] \quad (8)$$

$$\Delta E_N(T) = \frac{1}{Z_N} \left[ \Delta_N^{(+)} e^{-\beta \Delta_N^{(+)}} + \Delta_N^{(-)} e^{-\beta \Delta_N^{(-)}} \right] \quad (9)$$

$$Z_N = 1 + e^{-\beta \Delta_N^{(+)}} + e^{-\beta \Delta_N^{(-)}} \quad (10)$$

### Disorder averaging

We now consider a generic observable $O_N$, that only depends on the size $N$ of the cluster to which it is referred. The

average of the corresponding intensive quantity $o$ over the ensemble of Br-rich clusters can be generally written as

$$\langle o \rangle = \frac{1}{L^3}\,\langle O \rangle = \frac{1}{L^3}\sum_{N=1}^{\infty} O_N g_N \qquad (11)$$

where $g_N$ is the number of clusters of size $N$ present in the system. $g_N$ can be expressed as the probability of having a given cluster of size $N$ at a given location in space and with a given orientation and geometry, times the number of different locations, orientations and geometries that the cluster can take. Defining a cluster of size $N$ and delimited by $M$ sites, we can then write

$$g_N = \sum_M g_{NM} P_{NM}; \qquad (12)$$

here

$$P_{NM} \sim \exp(-x_0 N)\,\exp(-x_1 M) \qquad (13)$$

with $x_1 = |\log(1-2x)|$; $g_{NM}$ represents the number of possible geometrical realizations, rotations and translations of a cluster of size $N$ and delimited by $M$ sites. The translations are easily taken into account, by writing $g_{NM} = \alpha_{NM} L^3$. Therefore, we can generally write

$$g_N = A_N \exp(-x_0 N) L^3; \qquad (14)$$

where

$$A_N = \sum_M \alpha_{NM} \exp(-x_1 M) \qquad (15)$$

(we have reabsorbed the normalization of $P_{NM}$ into the coefficient $\alpha_{NM}$). It is clear that the exact knowledge of $A_N$ represents a challenging problem. Therefore we make the strong simplifying assumption that $A_N = A$ independent of $N$, and we use $A$ as a fitting parameter. (Though we assume $A$ is independent of $N$, it still depends on the doping concentration $x$.) Our local-gap model depends therefore on *two phenomenological parameters* only: $c$ and $A$. The above assumptions lead to the expression

$$\langle o \rangle \approx A \sum_{N=1}^{\infty} O_N \exp(-x_0 N) \qquad (16)$$

which can be cast in an integral approximation

$$\langle o \rangle \approx A \int_0^1 \frac{dy}{y^2}\, O(y)\,\exp(-x_0/y) \qquad (17)$$

justified by the fact that at low temperatures and low fields only large sizes contribute significantly to the thermodynamics of the system. The above integral can be easily computed numerically.

*Energy and specific heat in zero field*

We focus here on the thermodynamic properties in zero field, a limit in which the ensemble averages of the local gap model can be calculated almost exactly. The thermal energy (in units of $J_c$) of a cluster of size $N$ in zero field reads

$$\Delta E_N(t) = (c/N)/(1 + \exp(\beta c/N)/2) \qquad (18)$$

and the corresponding average thermal energy per spin reads

$$\Delta e(t) \approx 2cA \int_0^1 \frac{dy}{y}\,\frac{e^{-x_0/y}}{2 + e^{\beta c y}} \qquad (19)$$

A stationary phase approximation provides a very accurate expression for the above integral in the following form

$$\Delta e(t) \approx 2cA \left(\frac{\pi^2}{x_0 c}\right)^{1/4} t^{1/4} \exp(-2\sqrt{cx_0/t})\,. \qquad (20)$$

Consequently the specific heat reads

$$C(T)/k_B =$$
$$\frac{cA}{2}\left(\frac{\pi^2}{x_0 c}\right)^{1/4} t^{-3/4}\left(1 + 4\sqrt{cx_0/t}\right)\exp(-2\sqrt{cx_0/t}) \qquad (21)$$

giving Eq. (2) in the main text in the limit $t \to 0$.

*Comparison with QMC results*

Here we compare energy and magnetization obtained from the LGM model to the QMC results. The comparison is made between the *thermal energy* $\Delta E(T)$, and the *thermal magnetization* $\Delta m(T) = m(T) - m(0)$. This is more appropriate than comparing the simple energy and magnetization. Indeed the LGM cannot predict the value of the ground state energy, and it only assumes the form a scaling Ansatz on the local gap. Moreover, the $T = 0$ magnetization is not well captured by QMC at low fields, because it comes from extremely rare clusters whose size might be comparable with that of the simulation box, and whose probability to appear, $P_N$, is too small to be correctly sampled by the disorder averaging performed in the QMC simulation. In fact, at $T = 0$ the LGM model predicts the following expression for the magnetization:

$$m(h; T = 0) \approx \frac{A}{2x_0}\exp(-cx/h). \qquad (22)$$

This is the typical magnetization curve of the Bose glass [6, 7], which is exponentially suppressed at low fields. The essential singularity exhibited by the magnetization at $h = 0$ implies that the Mott glass phase, realized in zero field, has a vanishing linear susceptibility/compressibility, as well as vanishing susceptibilities to all orders of non-linearity, despite being gapless.

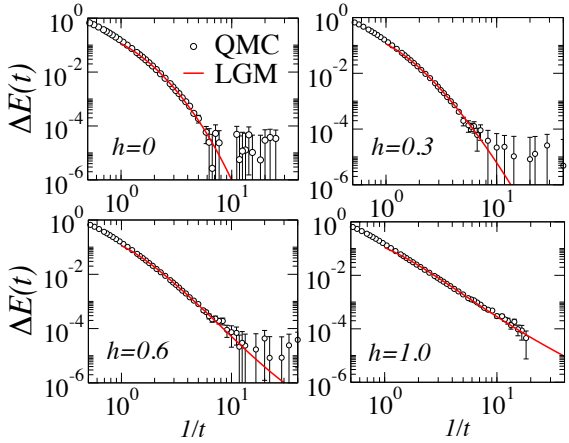

FIG. 4: Energy from LGM and QMC at various fields.

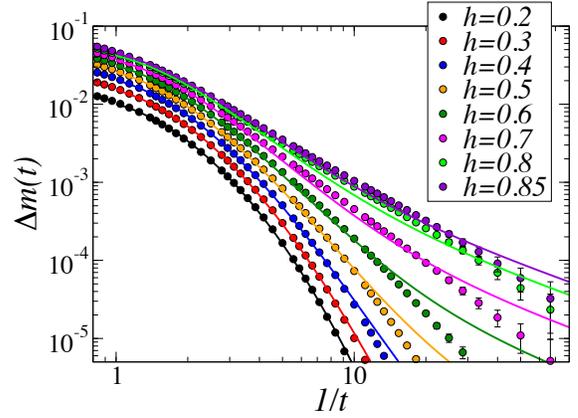

FIG. 5: Magnetization from QMC and LGM.

We compare the QMC data with the predictions of the LGM by first fitting the parameters $c$ and $A$ to the QMC data for the energy at $h = 0$. For the QMC parameters used in the simulation ($x = 7.5\%$, $J_c = 2.2$ K, $J_{ab} = 0.18$ K $J'_c = 2.35 J_c$, $D = 8.9$ K, $D' = D/2$), we find the best agreement between the LGM and the QMC data for $c = 3.75$ and $A = 4.55$. The $h = 0$ data for the energy are very well described by the LGM prediction. The values for $A$ and $c$ deviate from the ones that best fit the experimental results ($c = 3.02$ and $A = 2.69$) due to the strong sensitivity of the low-field thermodynamics to the actual value of the doping and of the Hamiltonian parameters. This sensitivity comes from the fact that the dominant thermodynamic response of the system in low fields is given by the Br-rich regions, as captured by the local-gap model, and that the probability distribution of the size of such regions is exponentially dependent on the doping value.

Remarkably, *without any other adjustable parameter*, we can extend the comparison between QMC and LGM to finite fields, finding an extremely good agreement up to fields $h \approx 1$ (corresponding to $H \approx 1.5$ T). The only significant deviations appear for $h \gtrsim h_{c1}$, due to the appearance of the finite-$T$ transition, which the LGM cannot account for. Moreover, the magnetization is also very well reproduced up to $h \approx 0.6$, again without any further adjustable parameter. We attribute the systematic deviation between the low-temperature QMC data and the LGM data to the fact that the low values of the magnetization ($\lesssim 10^{-4}$) are poorly sampled in the QMC simulations, due to the rarity of the regions with which they are associated, and to the finite size of the simulation boxes, which imposes an artificial upper cutoff on the size of the rare regions.

The fact that the LGM model captures the thermodynamics of the Hamiltonian model for Br-doped DTN so well makes it a very flexible tool for the comparison with the experiments. The microscopic details of the Hamiltonian are absorbed in

the two fitting parameters $c$ and $A$, while the doping and applied field are external parameters that can be varied continuously.

## ESTIMATE OF THE CRITICAL FIELDS AND $\phi$ EXPONENTS

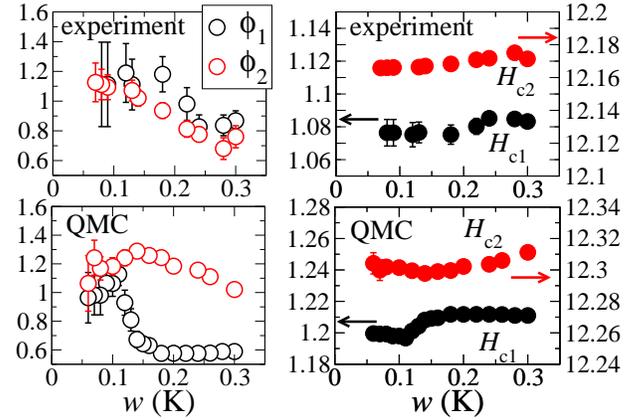

FIG. 6: Windowing estimate of the $\phi$ exponent and of the critical fields from experiments (AC susceptibility measurements) and QMC simulations.

The power-law scaling of the critical temperature as a function of the critical field, $T_c \sim |H - H_{c1}|^\phi$ and $T_c \sim |H - H_{c2}|^\phi$, holds only in the vicinity of the critical fields $H_{c1}$ and $H_{c2}$, and it can be considered as the asymptotic limit of the field-dependence of the critical temperatures in the low-temperature regime. Its validity is evident in the log-log plots in Fig. (4) of the main text. To extract $H_{c1}$, $H_{c2}$ and $\phi$ in the asymptotic limit one can adopt a *windowing* technique [8, 9],

which consists of approaching the low-temperature regime by fitting the experimental or numerical data over a temperature window $T \in [0, w]$ with a progressively smaller width. In practice we fit the temperature dependence of the critical field for the finite-temperature BEC, using the three parameter form $H(T) = H_{c1(2)} + AT^{1/\phi_{1(2)}}$ in the vicinity of the critical fields, for temperatures in the $[0, w]$ window. The results for both the exponents and critical fields are shown in Fig. 6. We see that both experiments and theory point towards a well-defined asymptotic regime with convergence of the critical fields and of the $\phi$ exponents towards a well-defined asymptotic value, reported in the main text. The value and error bar of the fitting parameters are chosen as those associated with the $w$ value at which the onset of convergence is observed. This allows us to exploit the largest number of data reproducing the supposed asymptotic limit.

---




\end{document}